%% file: main.tex
\documentclass[12pt,american]{article}
\usepackage[T1]{fontenc}
\usepackage[top=1.2in, bottom=1.2in, left=1.2in, right=1.2in]{geometry}
\usepackage{babel,amsmath,amsthm,amsfonts,graphicx,caption,subcaption,multirow,float,setspace,multirow,booktabs,collcell}
\usepackage{makecell}
\usepackage{xcolor}
\definecolor{darkblue}{RGB}{0,51,102}
\usepackage[colorlinks=true, citecolor=darkblue, linkcolor=darkblue, urlcolor=darkblue]{hyperref}
\usepackage[flushleft]{threeparttable}
\usepackage{tikz}
\usetikzlibrary{tikzmark}
\usetikzlibrary{decorations.pathreplacing}
\usepackage{newpxtext,newpxmath}

\newcommand\T{\rule{0pt}{2.6ex}}
\newcommand\Tnew{\rule{0pt}{3.6ex}}       

\newcolumntype{L}[1]{>{\raggedright\let\newline\\\arraybackslash\hspace{0pt}}m{#1}}
\newcolumntype{C}[1]{>{\centering\let\newline\\\arraybackslash\hspace{0pt}}m{#1}}
\newcolumntype{R}[1]{>{\raggedleft\let\newline\\\arraybackslash\hspace{0pt}}m{#1}}

\usepackage[authoryear,round]{natbib}
\usepackage{authblk}
\makeatletter
\newcommand{\specificthanks}[1]{\@fnsymbol{#1}}
\makeatother

\graphicspath{{./}}

\title{What Do City Codes Say About Climate Risk?\thanks{I thank Jenny Liu for excellent research assistance.}}

\author{Arianna Salazar-Miranda}
\affil{Yale School of the Environment}
\date{\today}

\begin{document}

{\let\newpage\relax\maketitle}
\vspace{-1em}

\begin{abstract}
\noindent Climate change is intensifying natural hazards, and preparing cities for them falls primarily to local governments. Yet what municipal codes say about hazards, and whether their requirements match each city's risk profile, has not been measured. I analyze 3.6 million sentences from 204 U.S. cities' municipal codes and compare each city's hazard text to its actual risk from FEMA's National Risk Index. I document four patterns. First, a small share of code text consists of provisions relevant for natural hazards (1.5\%), and codes in high-risk cities contain marginally more. Second, hazard coverage is highly concentrated on flooding (61\% of hazard-relevant text), even though flooding accounts for only 9\% of the average city's risk. Third, city codes emphasize emergency response over prevention. Emergency management is the largest regulatory category, and land use is the smallest. Fourth, codes track the types of risks relevant to each location, but only loosely. The median city would need to reallocate 70\% of its regulatory text to match its risk profile. The results suggest that U.S.\ city codes do not include regulation for most of the hazards cities face.

\end{abstract}
\clearpage
\section{Introduction}
\label{sec:intro}

Climate change is increasing the frequency and severity of natural hazards. The number of billion-dollar weather disasters in the United States has risen from three per year in the 1980s to more than twenty in recent years \citep{noaa_ncei}. Preparing cities for these hazards involves federal programs, state building codes, and local governments, but the enforceable rules that govern where and how development happens are ultimately assembled in each city's municipal code. These rules include zoning and land-use regulation, building codes, environmental ordinances, infrastructure requirements, and emergency management provisions. For example, a floodplain ordinance specifies how far structures must sit above the base flood elevation, a wildfire code requires vegetation management around buildings, and a coastal setback rule restricts development within a fixed distance of the shoreline. The text of a city's municipal code is therefore the main record of how the city requires development, construction, and emergency response given the hazards it faces.

Two strands of research have studied how cities prepare for natural hazards. The first examines regulation, focused mostly on floodplain ordinances under the National Flood Insurance Program (NFIP) \citep{kousky2014}. The second evaluates the content of adaptation and hazard mitigation plans \citep{wheeler2008, bierbaum2013, shi2015, brinkley_regalado_boswell2024}, finding that plans list many strategies but provide limited detail on how those strategies will be carried out \citep{berke2009, lyles2014, woodruff2016, han_laurian_dewald2021}. Codes record how individual cities respond to those hazards, and show how federal provisions are ultimately adopted and understood locally by municipalities. Yet what U.S.\ cities' codes say about the hazards they face, and whether their requirements match each city's risk profile, has not been measured.

Two recent developments make this measurement possible. First, large language models can read and classify millions of sentences of regulatory text at scale, producing structured measures of what each provision addresses and how. Such approaches have recently been applied to study zoning \citep{salazar_miranda_talen2025, mleczko_desmond2023, shanks2021} and to analyze the content of climate action plans \citep{bickel2017}, general plans \citep{brinkley_stahmer2024}, environmental justice plans \citep{brinkley_wagner2022}, and resilience plans \citep{fu_li_zhai2023}. Second, FEMA's National Risk Index, released in 2020, scores every U.S.\ census tract across 18 distinct hazard types, producing a comparable hazard profile for cities. Together they allow each city's code to be compared directly to a standardized profile of the hazards it faces. I use these data to ask three questions. What hazards do U.S. cities regulate in their codes? Which regulatory mechanisms do they employ to do so? And do they regulate the hazards they actually face?

To answer these questions, the full text of municipal codes for 204 U.S. cities is collected from Municode and American Legal Publishing, the two largest online repositories of municipal legislation, yielding 3.6 million sentences. Each sentence is then classified with GABRIEL, an open-source library that uses large language models to score attributes in text \citep{asirvatham2026}. Every sentence is scored on a 0-100 scale for its relevance to each of the 18 natural hazard types in FEMA's National Risk Index (NRI). Each hazard-relevant sentence is also assigned to one of three regulatory mechanism categories: land-use and environmental regulation, building codes and infrastructure, or emergency management. These sentence-level scores are aggregated to the city level and paired with NRI risk shares to compare, for each city, its hazard text profile to its hazard risk profile.

Using these data, I document four facts. First, city codes devote a small share of their text to natural hazards, and codes in high-risk cities devote marginally more.

Second, hazard coverage is highly concentrated on flooding, which dominates hazard-relevant text even though it accounts for only a small share of the average city's risk. Hazards that cause substantial damage and mortality, including heat waves, hail, and tornadoes, account for little text in city codes.

Third, city codes emphasize emergency response over prevention. Of the three mechanisms, emergency management, which governs what a city does once a hazard occurs, is the largest category, and land-use restrictions, which keep development out of hazardous locations in the first place, the smallest. Which mechanism a city uses to regulate a hazard also depends on the hazard itself. Landslides are regulated mainly through land use, earthquakes through building codes, and tornadoes through emergency management. Emergency management provisions are also the most likely to be written in mandatory language, and land-use provisions the least.

Fourth, codes track risk only loosely, and do not include regulation for most of the hazards cities face.

\section{Measurement}
\label{sec:measurement}

\subsection{Data and classification}
\label{sec:data-classification}

Municipal codes are collected from Municode and American Legal Publishing, the two largest online repositories of municipal legislation in the United States. The sample consists of 204 cities, all with populations of at least 100,000. Of these, 174 publish on Municode and 30 on American Legal Publishing, which is where most of the remaining large cities publish. I collect from both because several of the largest U.S. cities, among them Los Angeles, Chicago, Philadelphia, and San Francisco, are absent from Municode.

I use a threshold of 100,000 because cities of this size share a broadly similar governance scale in the sense that they have their own planning, building, and emergency management departments. The 204 cities are home to 61 million residents, 62\% of the population living in U.S. cities above this threshold. The sample spans 42 states and all four U.S. Census regions (87 South, 72 West, 33 Midwest, 12 Northeast; see Supplementary Figure~\ref{fig:appendix-state-coverage} and Supplementary Table~\ref{tab:appendix-sample-summary} for the full list). Those four regions hold 38\%, 24\%, 21\%, and 17\% of the U.S. population.

The full text of each city's code is downloaded as of 2023 (Municode) or 2026 (AmLegal) and split into individual sentences, yielding 3.6 million sentences. I classify at the sentence level because the sentence is the smallest unit consistently defined across cities' varied formatting. Both platforms publish a single up-to-date master text for each city, integrating amendments as they are adopted, so each city contributes one current version of its code.

Each sentence is classified using GABRIEL, an open-source library that prompts a large language model to score attributes in text \citep{asirvatham2026}. The underlying model is GPT-4o-mini. Before scoring, a simple keyword filter removes sentences clearly unrelated to natural hazards, using a dictionary of hazard-related terms (see Methods). For each remaining sentence the model returns three main outputs. The first is the sentence's overall relevance to natural hazards. The second is its relevance to each of the 18 hazard types defined by the NRI, which are listed in Supplementary Table~\ref{tab:appendix-keywords}. The third is the sentence's enforceability. 

Hazard relevance is a score from 0 to 100 reflecting how directly a sentence addresses natural hazards in general. The model returns 100 for sentences that explicitly state a hazard-related requirement, such as elevating a structure above the base flood elevation, and 0 for sentences with no hazard content. I treat a sentence as hazard-relevant if its overall hazard relevance score, or any of its 18 hazard-specific scores, exceeds 50. A score of 50 means the model considers the sentence to meaningfully address the hazard. Because few sentences score in the 40 to 60 range, the exact cutoff within this range has little effect on which sentences are classified as hazard-relevant (Appendix Figure~\ref{fig:appendix-relevance-distribution}).

For each of the 18 NRI hazards, the model also returns a separate score from 0 to 100 reflecting how directly a sentence addresses that hazard. A single sentence can score above 50 on more than one hazard. For example, a stormwater drainage provision may score high on both riverine flooding and strong wind. Each sentence is assigned fractionally across the relevant hazards, in proportion to its scores, and these contributions are summed across all sentences in a city's code. This yields the share of hazard-relevant text devoted to each of the 18 hazards, which I treat as the city's hazard text profile.

Enforceability is a score from 0 to 100 reflecting how binding a sentence is. The model returns 100 for mandatory language, such as \emph{shall}, \emph{must}, or \emph{required}, and 0 for aspirational language, such as \emph{should}, \emph{may}, or \emph{encourage}. I use this score to ask whether the mechanisms cities write most about are also the ones where their rules are most binding.

To validate the hazard-relevance classification, 299 sentences were coded by hand, independently of the model's scores. The sample covers sentences classified by the model as hazard-relevant, sentences it classified as not hazard-relevant, and sentences the keyword filter removed before the model saw them. Of the sentences classified as hazard-relevant, 88\% were hazard-relevant to the human coder, and the model identified 86\% of the sentences the human coder marked as hazard-relevant. Of the 89 sentences the keyword filter had removed, 2 were hazard-relevant, so the filter discards almost no hazard-relevant text (Supplementary Table~\ref{tab:appendix-validation}).

These sentences do not all date from the same period, because cities amend their codes piecemeal rather than rewriting them. I date each section from its ordinance history. The median hazard-relevant sentence was adopted in 1998 and last amended in 2020, and only 4\% were last amended before 2000. Most hazard provisions are therefore fairly recent (Supplementary Table~\ref{tab:appendix-code-vintage}).

To measure each city's actual hazard exposure I use the underlying risk scores from FEMA's NRI. The NRI reports a risk percentile from 0 to 100 for each census tract and each of the 18 hazards. Scores are aggregated from tracts to cities by averaging across all census tracts whose area falls at least 50\% within the city's boundary, then normalized across hazards so that each city's 18 scores sum to one. The result is the city's hazard risk profile, an 18-element vector in which each element represents the share of the city's total hazard risk attributable to that hazard. I compare this risk-based measure with the text-based measure to assess whether the hazards a city regulates match the hazards it actually faces.

\section{Results}
\label{sec:results}

\subsection{What hazards do U.S.\ cities regulate?}
\label{sec:results-hazards}

\begin{figure}[h!]
\centering
\includegraphics[width=\textwidth]{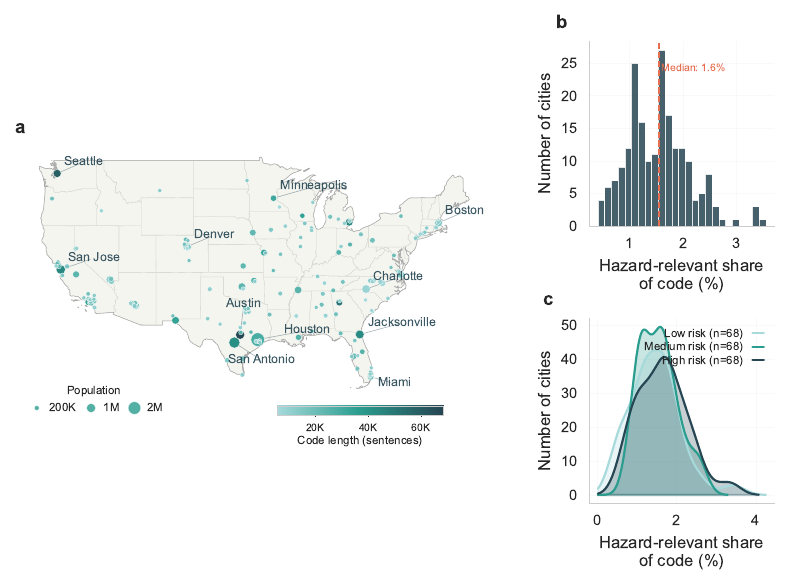}
\caption{\textbf{Sample of 204 U.S.\ cities and the hazard-relevant share of municipal code.} (a) Map of the sample. Point size scales with population; color scales with total sentence count in each city's municipal code. (b) Distribution across cities of the share of municipal code text classified as hazard-relevant. (c) Kernel density of the hazard-relevant share across cities, separately by NRI composite risk tercile.}
\label{fig:sample}
\end{figure}

I begin by asking how much of the municipal code of U.S.\ cities is devoted to natural hazards. Figure~\ref{fig:sample}A maps the 204 cities in the sample. The sample spans the continental United States, includes both coastal and inland cities, and contains substantial variation in population (roughly 100,000 to 4 million) and code length (about 5,500 to 68,500 sentences per city).

Figure~\ref{fig:sample}B plots the distribution of hazard-relevant text across cities. The median city devotes 1.5\% of its code to natural hazards, about 250 sentences. At the 90th percentile, cities devote about 2.3\%, and the cities that allocate the largest share (Columbia SC, North Charleston, Jackson) reach about 3.5\%. At the bottom of the distribution, Boston, Centennial, and Syracuse allocate less than 0.5\%; Boston's code contains 62 hazard-relevant sentences.

If cities regulate the hazards they face, a natural expectation is that cities at greater risk devote more of their code to hazards. Figure~\ref{fig:sample}C plots the distribution of hazard-relevant share separately for cities in the low, medium, and high terciles of FEMA's NRI composite risk score. Higher-risk cities do allocate more, but the difference is small. A one-standard-deviation increase in a city's total NRI risk is associated with a 0.15 percentage point higher share of hazard-relevant text ($p<0.001$, $R^2=0.06$; Supplementary Table~\ref{tab:appendix-risk-regression}). Because bigger cities tend to have longer codes (Supplementary Figure~\ref{fig:appendix-pop-code-length}), I also control for total code length. The risk coefficient is similar (0.10 pp, $p=0.008$). Even the high-risk tercile stays below the 2.3\% observed at the 90th percentile.

\begin{figure}[h!]
\centering
\includegraphics[width=\textwidth]{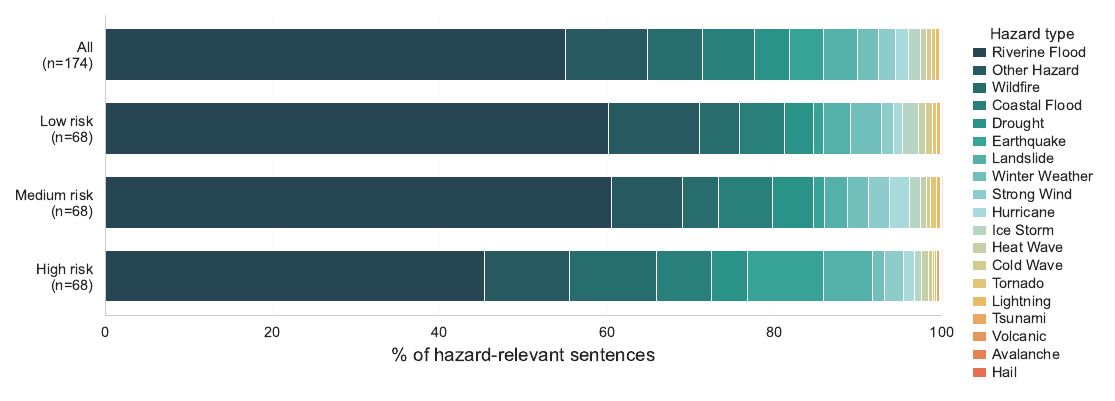}
\caption{\textbf{Share of hazard-relevant text devoted to each of the 18 NRI hazards.} Stacked bars show the share of hazard-relevant sentences in each hazard category, across the full sample of 204 cities and within each NRI composite risk tercile.}
\label{fig:concentration}
\end{figure}

I now look at the composition of hazard regulation across the 18 NRI hazards. Figure~\ref{fig:concentration} plots the share of hazard-relevant text devoted to each of the 18 NRI hazards, across the full sample and in each NRI composite risk tercile. The figure shows that riverine flooding dominates across every grouping, accounting for 61\% of hazard-relevant text across the full sample. This concentration is consistent with the federal regulatory infrastructure built around flooding, which requires participating cities to adopt minimum floodplain ordinances under the National Flood Insurance Program \citep{kousky2014}.

The share devoted to riverine flooding is smaller in higher-risk cities, falling from 65\% in the low-risk tercile to 53\% in the high-risk tercile ($p<0.01$). The difference is reallocated to other hazards. Wildfire rises from 6\% to 11\% ($p<0.001$) and earthquake from 1.5\% to 9\% ($p<0.001$). Higher-risk cities therefore address a broader set of hazards than lower-risk ones. Cities adjust the composition of their regulation to the hazards they face, even when they do not write more hazard-relevant text overall. Even so, hail, tornado, lightning, and cold wave each account for less than 1\% of hazard-relevant text in every risk tercile.

\subsection{What regulatory mechanisms do cities employ?}
\label{sec:results-mechanisms}

I now ask which mechanisms cities use to regulate hazards. I divide mechanisms into three categories: land use, building regulation, and emergency management. Land use governs where development is allowed and includes rules such as zoning, stormwater management, and wetland protections. Building regulation governs how development is built and includes construction standards, infrastructure requirements, and drainage systems. Emergency management governs how the city responds to an event and includes evacuation plans, shelters, warning and monitoring systems, and provisions for recovery and reconstruction. For example, a sentence requiring structures to sit above the base flood elevation is classified as building regulation, whereas a sentence directing the mayor to declare a state of emergency is classified as emergency management. How a city allocates its regulation across these three mechanisms therefore determines whether it prevents hazard exposure, reduces damage when exposed, or only manages the aftermath.

\begin{figure}[h!]
\centering
\includegraphics[width=\textwidth]{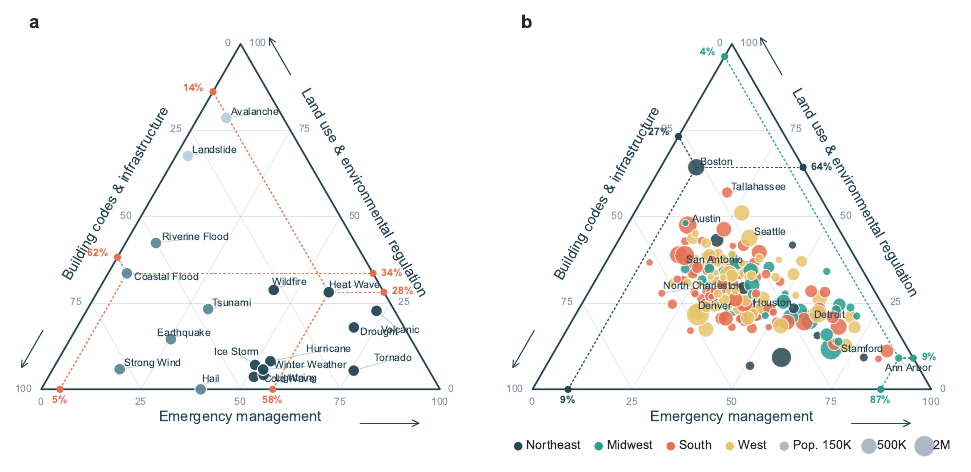}
\caption{\textbf{Regulatory mechanism mix across hazards and across cities.} Each point is positioned in the triangle by its share of hazard-relevant text in land use and environmental regulation, building codes and infrastructure, and emergency management. (a) One point per hazard; color shade encodes the dominant category (light = land use, medium = building codes, dark = emergency management). Orange reading guides for Coastal Flood and Heat Wave illustrate how to read the percentages off each axis. (b) One point per city; color shows U.S.\ Census region and point size scales with population. Reading guides for Boston and Ann Arbor mark the two extremes of the city-level distribution.}
\label{fig:mechanisms}
\end{figure}

Figure~\ref{fig:mechanisms}A plots each of the 18 hazards as a point in a triangle whose three corners correspond to the three mechanism categories. The closer a hazard sits to a corner, the larger the share of its regulatory text in that mechanism.

Three groups of hazards stand out in the triangle. Landslide and avalanche sit near the land-use corner. Of all rules addressing landslide, 68\% are land-use rules. For avalanche, the share is 79\%. A natural explanation is that these hazards occur in locations that can be mapped in advance, like steep slopes prone to mass movement and snow-prone mountainsides, which lets cities zone development away from those areas.

Coastal flood, strong wind, and earthquake sit near the building-codes corner. Of all rules addressing coastal flood, 62\% are building-code rules. For strong wind the share is 77\%, and for earthquake 60\%. One plausible interpretation is that these hazards can be addressed through structural design standards such as seismic anchorage, wind-load requirements, and elevation above flood level, which fit naturally within building codes.

Tornado, drought, heat wave, and hurricane sit near the emergency corner, the most populated of the three. Of all rules addressing tornado, 76\% are emergency-management rules. For drought the share is 69\%, for heat wave 58\%, and for hurricane 53\%. Cities may use emergency rules because these hazards cannot be prevented through zoning or reduced through engineering, leaving only response measures like warning systems, evacuation, and emergency shelter.

Figure~\ref{fig:mechanisms}B shows the same classification at the city level. Each point is now one city, positioned by its allocation across the three mechanisms. The mean city devotes 39\% of its hazard-relevant text to emergency management, more than to building codes (32\%) or land use (29\%), so cities' codes devote more text to response than to prevention. In counts, that is 80 sentences on emergency management against 75 on building codes and 67 on land use. Emergency management is the largest of the three categories in 93 of the 204 cities, against 72 for building codes and 35 for land use, with the remaining 4 tied. Emergency is the largest category in every U.S. Census region, though it accounts for a larger share in the Midwest and Northeast (44--46\%) than in the South and West (37\%).

Codes differ in how they balance the three mechanisms. In Boston, 64\% of hazard-relevant text is in land-use rules. In Ann Arbor, 87\% is in emergency management. In San Jose, the text is divided roughly evenly across the three.

\begin{figure}[h!]
\centering
\includegraphics[width=0.7\textwidth]{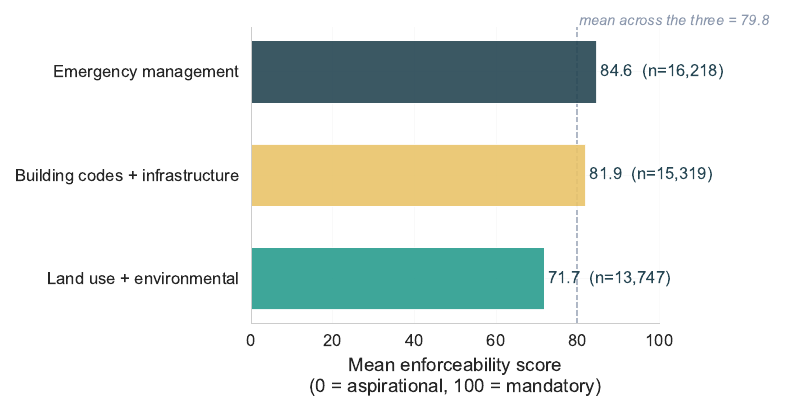}
\caption{\textbf{Mean enforceability by regulatory mechanism.} Bars show the mean enforceability score of hazard-relevant sentences in each of the three mechanism groups, on a scale from 0 for aspirational language to 100 for mandatory language. The dashed line marks the mean across the three groups combined. $n$ is the number of sentences in each group (see Methods).}
\label{fig:enforceability}
\end{figure}

More hazard-relevant text devoted to a mechanism does not necessarily mean its rules are more binding. Figure~\ref{fig:enforceability} plots the mean enforceability of hazard-relevant sentences in each of the three mechanism categories, on a scale running from aspirational to mandatory language. Emergency management scores 84.6, building codes and infrastructure 81.9, and land use and environmental regulation 71.7, against a mean of 79.8 across the three mechanisms combined. All three differences are significant at $p<0.001$. Emergency management has both the largest share of hazard-relevant text and the highest mean enforceability, and land use the lowest of each.

\subsection{Do cities regulate the hazards they actually face?}
\label{sec:results-alignment}

One way to compare cities is whether their regulation addresses the hazards they actually face. A city facing high flood risk matches its risk profile if it devotes more of its code to flooding than to other hazards, and differs from it if its text is concentrated elsewhere. I now compare each city's hazard risk profile to its hazard text profile. The risk profile is the share of NRI risk attributable to each hazard. The text profile is the share of hazard-relevant text in each of the 18 hazards.

\begin{figure}[h!]
\centering
\includegraphics[width=\textwidth]{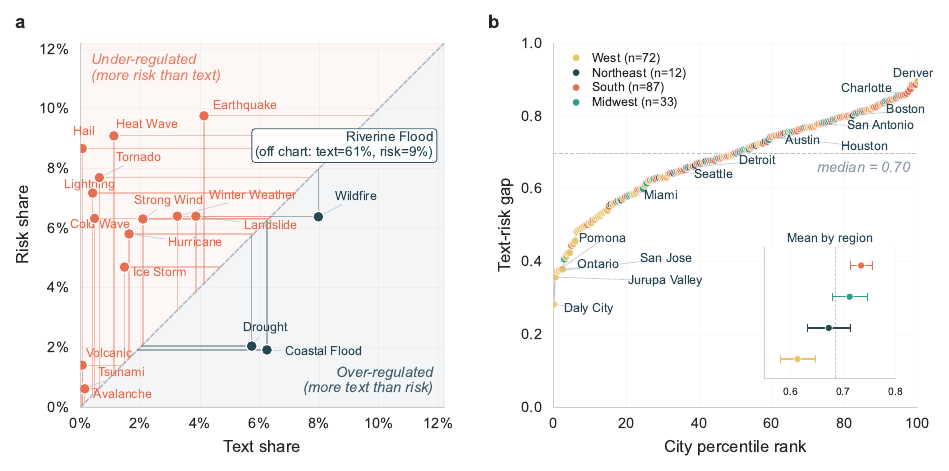}
\caption{\textbf{Alignment between regulatory text and actual hazard risk.} (a) Per-hazard text share vs.\ risk share, averaged across the 204 cities. Each dot is one hazard. The 45-degree line marks perfect alignment. Hazards above the line feature more prominently in cities' risk profiles than in their codes. Hazards below feature more prominently in the codes. Riverine flooding is annotated off-chart because its text share dwarfs every other hazard. (b) City-level distribution of the text-risk gap. For each city, the gap equals half the sum of absolute differences between its text profile and its risk profile across the 18 hazards. A value of 0 means perfect alignment; 1 means complete mismatch. Cities are ordered along the x-axis by their gap. The inset shows the regional means with 95\% confidence intervals.}
\label{fig:alignment}
\end{figure}

Figure~\ref{fig:alignment}A plots the mean text share against the mean risk share for each of the 18 hazards. Each dot is one hazard, averaged across the 204 cities. The 45-degree line marks perfect alignment. Hazards above the line feature more prominently in cities' risk profiles than in their codes. Hazards below the line feature more prominently in the codes than in the risk profiles. Riverine flooding sits far below the line, accounting for 61\% of the average city's hazard-relevant text but only 9\% of its average risk. Heat waves, hail, and tornadoes lie far above the line, each contributing 8 to 9\% of average risk but 1\% or less of regulatory text. Riverine flooding takes the largest share of text relative to its risk share, followed by coastal flooding, drought, and wildfire. Every other hazard takes a smaller share of text than of risk.

Figure~\ref{fig:alignment}B shows, for each of the 204 cities, the share of hazard-relevant text that would need to be reallocated across hazards to match the city's risk profile. I call this the text-risk gap. A value of 0 means perfect alignment; 1 means complete mismatch. The median city's gap is 0.70. The smallest, 0.28, is Daly City's. The largest, closer to 0.90, are Denver, Killeen, and Jackson. Western cities have smaller gaps on average (mean 0.61) than cities in the Northeast (0.67), Midwest (0.71), or South (0.73).

The gap compares a city's full text profile with its full risk profile. Restricting the same comparison to the hazards that pose the largest risks, in three quarters of cities at least one of the three largest receives the equivalent of at most one sentence. In 30\% of cities this is true of the single largest risk (Supplementary Figure~\ref{fig:appendix-silence}). This is most common for hail, lightning, cold wave, and tornado, the same hazards that sit farthest above the 45-degree line in Figure~\ref{fig:alignment}A.

\subsection{Robustness}
\label{sec:results-robustness}

The findings are similar under alternative choices about how the sample is selected and how text is counted. The main results are comparable when recomputed on cities above higher population thresholds (Supplementary Table~\ref{tab:appendix-threshold}), and when each city is reweighted so its Census region contributes in proportion to that region's share of the U.S. population (Supplementary Table~\ref{tab:appendix-region-weighted}). Counting by sentence rather than by word could matter if provisions for some hazards are drafted in longer sentences than others. Weighting each sentence by its word count instead gives near-identical per-hazard rankings (Spearman rank correlation 0.99, mean absolute difference in per-hazard share 0.15 percentage points; Supplementary Figure~\ref{fig:appendix-wordweight}).

Another concern is that what a city's code says about hazards reflects federal requirements or language copied from elsewhere rather than decisions the city made. The NFIP floodplain requirement noted above is one such source, and cities might also copy regulatory language from each other or from model codes. If either were widespread, the concentration on flooding, the emphasis on response, and the size of the text-risk gap would say more about federal mandates and shared templates than about how cities choose to regulate.

I can test both possibilities directly from the text. First, I identify sentences that explicitly reference external frameworks (FEMA, NFIP, state codes, and similar) using a dictionary of 384 named references and acronyms built from the corpus itself, the most frequently cited of which are listed in Supplementary Table~\ref{tab:appendix-external-frameworks}. Sentences without such references are counted as text without external references. Second, I identify sentences that appear verbatim in five or more cities, capturing language that cities have copied from each other or from model codes.

Across the sample, 86\% of hazard-relevant text does not explicitly cite federal or state frameworks, and only 4\% of sentences are boilerplate (Appendix Figure~\ref{fig:appendix-local-boilerplate}). Most hazard-relevant text is therefore written without direct citation of external frameworks and is not verbatim copy from other cities.

The remaining 14\% is informative in its own right, because a sentence that names a framework is a sentence whose external influence can be traced. That influence is concentrated on flooding. Riverine and coastal flooding provisions cite an external framework in 20\% and 22\% of sentences, against 8\% for wildfire, 2\% for drought, and 1\% for heat wave.

\section{Discussion}
\label{sec:discussion}

As climate change intensifies natural hazards, how cities adapt depends on the binding rules they have in place. In the United States, those rules sit in municipal codes. The findings show that municipal codes devote little text to hazards, concentrate that text on flooding, and allocate more of it to response than to prevention. Across cities, what codes regulate only loosely tracks the hazards each city faces.

Two features of U.S.\ hazard governance help interpret these patterns. First, the dominance of flooding in hazard-relevant text is consistent with the federal architecture around flood risk. The National Flood Insurance Program conditions federal disaster assistance on participating cities adopting minimum floodplain-management ordinances, and its Community Rating System reduces insurance premiums for cities that go beyond the minimum. The minimum area these ordinances must cover is the Special Flood Hazard Area shown on FEMA's flood maps, which FEMA draws and revises through a process open to challenge by communities and property owners \citep{pralle2019}. No comparable federal mandate exists for any other hazard, which helps explain why the same intensity of local regulation does not appear elsewhere. Flooding is also the hazard cities have been regulating longest, since settlement has always clustered near water. Only 10\% of the world's population lives more than 10 kilometres from a surface freshwater body \citep{kummu2011}. Second, the balance between response and prevention reflects a long-standing feature of U.S.\ urban development. Restrictive zoning and land-use controls, the most direct preventive tools, are politically costly to adopt because they tend to lower property values, slow growth, and generate concentrated opposition from landowners and developers \citep{glaeser_gyourko2018, einstein_glick_palmer2019}. Cities may therefore favor response measures even where preventive ones would act more directly on exposure.

The analysis has several limitations. First, municipal codes are the binding form of regulation, but cities also operate through informal practices and plans that work alongside the codes. A city with little hazard-relevant text in its code may still address those hazards through these other channels, which the analysis cannot directly observe. Second, the NRI is one of several available measures of hazard risk. Alternative sources, such as modeled climate projections, might rank hazards differently for some cities. Third, the sample is restricted to cities with population of at least 100,000 that publish their codes on Municode or American Legal Publishing. Smaller cities are not represented and may follow different regulatory practices than I can characterize here. Fourth, I cannot say whether codes with a large text-risk gap actually leave cities less protected. The analysis measures what is written, not what gets enforced or what its effects are in practice. Fifth, the building-code category includes some requirements that originate in model codes. There is no federal building code, but states adopt some version of the International Building Code and International Residential Code, varying in how they amend them and in how much local variation they permit. Text in this category may therefore reflect a state's adoption decisions as much as a city's own. 

More broadly, this paper provides a baseline measurement of how U.S.\ cities have written their codes around natural hazards. By classifying regulatory text at scale and comparing it to standardized risk profiles, the framework shows how attention to specific hazards compares to the risks cities face. As cities revise their codes and as the hazard landscape itself evolves with climate change, the framework can be reapplied to track how regulation changes relative to the hazards it addresses.

\clearpage
\section*{Methods}
\label{sec:methods}

The analysis combines three main steps. First, for each city I build a hazard text profile that captures what its municipal code says about each of the 18 NRI hazards. Second, for each city I build a hazard risk profile that captures how much risk it faces from each of those same hazards. Third, I combine the two profiles into a text-risk gap that summarizes how far a city's regulatory attention is from its risk exposure. I also use the classified text to test whether the patterns I observe reflect independently drafted rules or text citing external frameworks. The following subsections describe each step.

\subsection*{Hazard text profile}

The hazard text profile for each city captures what the city's municipal code says about each of the 18 NRI hazards. I build it by classifying every sentence in the code, then aggregating to the city level. Classification proceeds in two parts. A keyword pre-filter first removes sentences obviously unrelated to hazards, and GABRIEL then scores the remaining sentences on hazard relevance and the specific hazards each addresses.

\paragraph{Keyword pre-filter.} To avoid running the language model on millions of sentences unrelated to hazards, I first apply a simple keyword pre-filter. Each sentence is checked against a dictionary of hazard-related terms organized by hazard type. For each of the 18 NRI hazards, the dictionary includes hazard-specific terms (such as ``earthquake'', ``seismic'', ``fault line'', and ``liquefaction'' for earthquake) and a broader set of general hazard-related terms (such as ``hazard'', ``disaster'', ``emergency management'', ``evacuation'', and ``mitigation''). The full list of terms is given in Supplementary Table~\ref{tab:appendix-keywords}. Sentences matching at least one term are sent to GABRIEL for scoring. A fixed dictionary cannot anticipate every way a code refers to a hazard, so I also ask a language model to read the sentences the keyword filter rejected and flag any that use hazard-related engineering terms the dictionary misses, such as ``impact-rated glazing'' or ``unreinforced masonry''. Flagged sentences join the pool sent to GABRIEL. Sentences that neither the keyword filter nor the LLM check flag are assigned a score of zero on all attributes. The keyword list is intentionally broad to minimize false negatives, since false positives are caught by GABRIEL's low scores at the next stage. Of the 3.6 million sentences in the corpus, 300,979 are sent to GABRIEL for scoring (169,371 through the keyword filter and an additional 131,608 through the LLM check). To check that the pre-filter does not discard hazard-relevant content, I manually inspected a random sample of 89 dropped sentences. Only 2.2\% (Wilson 95\% CI: 0.6\%--7.8\%) would have been called hazard-relevant by a human, indicating that the pre-filter discards almost no hazard-relevant text (Supplementary Table~\ref{tab:appendix-validation}, Panel II).

\paragraph{Classification using GABRIEL.} For each sentence that passes the pre-filter, I use GABRIEL to score it on hazard relevance and the specific hazards it addresses. I run this classification in two stages so that the more expensive 18-hazard scoring is applied only to sentences that are at least plausibly hazard-relevant. In the first stage, every sentence that passes the keyword pre-filter is scored on overall hazard relevance. Sentences with a Stage 1 hazard relevance score of at least 15 proceed to the second stage, in which they are scored on the 18 hazard-specific relevance scores and on the regulatory mechanism category. The Stage 2 threshold of 15 sits well below the hazard-relevant cutoff of 50 used in the analysis, ensuring that any sentence that could plausibly be hazard-relevant receives the full classification. Of the 300,979 sentences scored by GABRIEL, 67,382 clear Stage 1's threshold of 15 and receive the full 18-hazard classification, and 52,154 clear the hazard-relevance threshold of 50 used in the analysis.

For each attribute, GABRIEL prompts the model with the sentence and a short description of what the attribute means, and the model returns a score from 0 to 100. For the earthquake attribute, for example, the description tells the model that high scores correspond to sentences explicitly about earthquakes, seismic risk, or earthquake-resistant construction. In Stage 1 the model also scores each sentence on an enforceability attribute from 0 (aspirational language: ``should'', ``may'', ``encourage'') to 100 (mandatory language: ``shall'', ``must'', ``required''), which I use to check whether the mechanisms cities write most about are also the ones where their rules are most binding. The exact text of each attribute definition passed to the model is reproduced in Supplementary Table~\ref{tab:appendix-gabriel-prompts}.

When I report mean enforceability by regulatory mechanism, I average the score within each of the three mechanism groups and exclude the 6{,}869 sentences the model scores 0 on the mechanism attribute, which marks text that is not a regulation. The three groups contain 16{,}218, 15{,}319, and 13{,}747 of the 52{,}154 hazard-relevant sentences, and their mean enforceability is 79.8. Including the excluded sentences would lower it to 73.3, since text that is not a regulation averages 30.5. To test whether the groups differ, I compute each city's mean enforceability within each group and compare the groups pairwise with paired $t$-tests across the 204 cities.

Prior work reports variable agreement between language-model and human coding in planning research \citep{fu_wang_li2024}. See \citet{fu_etal2025} for a wider discussion of language models in planning. To validate the classifications, 299 sentences spanning 135 cities were coded by hand, without reference to how the model had scored them. They were drawn at random from three groups, 148 that GABRIEL classified as hazard-relevant, 62 that it classified as not hazard-relevant, and 89 that the keyword filter removed before GABRIEL saw them. On the 210 sentences the model rated, precision, recall, and F1 for hazard-relevance are 0.88, 0.86, and 0.87 (Supplementary Table~\ref{tab:appendix-validation}, Panel I). Per-hazard F1 ranges from 0.33 (heat wave, $n{=}5$) to 1.00 (hail, $n{=}8$). The sample also covers all 18 hazards evenly, between 4 and 12 sentences each. For example, riverine flooding contributed 8 sentences even though it accounts for 61\% of hazard-relevant text, and hail contributed 8 as well.

\paragraph{Vintage of the sections containing hazard provisions.} To date the hazard provisions, I use the ordinance history printed at the end of each code section, such as ``(Ord. No. 6300, \S\ 1, eff. 3-4-86)'' or ``(Ord. 466-C-S, passed 9-25-80; Am. Ord. 1082-C-S, passed 11-28-06)''. I read the year in all three of the forms publishers use, a four-digit year, a two-digit year inside a month-day-year date, or a year inside the ordinance number itself. I take the earliest year in a section as its date of first adoption and the most recent as its date of last amendment, and give both dates to every hazard-relevant sentence in that section. Dates are available for 39{,}642 of the 52{,}154 hazard-relevant sentences (76\%). The median hazard-relevant sentence sits in a section first adopted in 1998 and last amended in 2020, and only 4\% sit in a section last amended before 2000 (Supplementary Table~\ref{tab:appendix-code-vintage}).

\paragraph{Text devoted to a city's largest risks.} I rank the 18 hazards within each city by that city's risk profile, so its largest risk is the hazard contributing the greatest share of its total NRI risk. I then count the sentences each hazard receives in that city. A sentence addressing more than one hazard is assigned fractionally across them, so these counts are fractional, and I record the hazards receiving the equivalent of at most one sentence. Counting two sentences instead of one, the largest risk falls below the count in 40\% of cities and at least one of the three largest does in 82\%. Counting three, the two figures are 50\% and 87\%.

\paragraph{Robustness to sentence-level counting.} One alternative to counting sentences is to weight each sentence by its number of words, so that a hazard's share reflects the words devoted to it. I therefore recompute each city's text profile weighting every sentence by its word count rather than counting each equally. Supplementary Figure~\ref{fig:appendix-wordweight} compares the two, and the result is reported with the other robustness checks in the Results.

\subsection*{Hazard risk profile}

To measure each city's actual hazard exposure independently of its code, I use FEMA's National Risk Index. The NRI reports a risk percentile from 0 to 100 for each U.S.\ census tract and each of the 18 hazards. Scores are aggregated from tracts to cities by averaging across all census tracts whose area falls at least 50\% within the city's boundary, then normalized so that each city's 18 hazard scores sum to one, yielding the city's hazard risk profile. The risk profile is constructed entirely from observed hazard data and is independent of the municipal code text, which allows the risks cities regulate to be compared with the risks they actually face.

\subsection*{Text-risk gap}

The text-risk gap summarizes how far the city's regulatory attention is from its hazard exposure. I compute it as the total variation distance between the city's hazard text profile and its hazard risk profile across the 18 hazards. Both profiles are 18-element vectors that are non-negative and sum to one, so they can be treated as probability distributions over hazards.

Letting $t_{c,h}$ denote city $c$'s text share for hazard $h$ and $r_{c,h}$ its risk share for the same hazard, the text-risk gap is
\[
\text{gap}_c = \tfrac{1}{2}\sum_{h=1}^{18} |t_{c,h} - r_{c,h}|.
\]
Each term in the sum is the absolute difference between the city's text share and its risk share for one hazard, which captures how far the city's regulatory attention to that hazard deviates from the hazard's share of its total risk. Halving the sum puts the gap on a 0 to 1 scale.

The score takes values in $[0, 1]$. A value of 0 means the text profile matches the risk profile exactly. A value of 1 means the two profiles are completely disjoint. The score can be interpreted directly as the share of hazard-relevant text that would need to be reallocated across hazards to bring the text profile into alignment with the risk profile.

\subsection*{External framework references}

A natural concern is that the patterns I observe in cities' codes do not reflect local choices but external constraints. Federal regulation could be one source, as could language copied verbatim across cities. To rule these out, I apply two complementary detection procedures and report the share of hazard-relevant text in each category.

To identify sentences referencing external regulatory frameworks, I build a dictionary of such frameworks from the corpus itself. First, all 3.6 million sentences are scanned for references to external regulations, such as ``International Building Code,'' ``National Flood Insurance Program,'' state building and fire codes (e.g., ``California Building Code''), and Code of Federal Regulations citations (e.g., ``44 CFR Part 60''). Second, I keep only frameworks that appear in at least three cities and whose name contains a hazard- or built-environment-related keyword (such as ``building'', ``flood'', ``fire'', ``coastal'', or ``seismic''). The three-city threshold filters out idiosyncratic one-off references, and the keyword filter restricts the dictionary to frameworks relevant to hazard regulation. The resulting list is the dictionary of external regulatory frameworks; it contains 384 entries in total, comprising 142 named frameworks (e.g., ``International Building Code,'' ``National Flood Insurance Program'') and 242 acronyms (e.g., ``FEMA,'' ``NFPA,'' ``BFE''). The 25 most frequently cited entries are listed in Supplementary Table~\ref{tab:appendix-external-frameworks}. Any sentence containing at least one entry from this dictionary is then flagged as referencing an external framework. Sentences without such references are counted as text without external references.

For boilerplate detection, I identify sentences that appear verbatim in five or more cities. Sentences are normalized before matching by lowercasing the text and collapsing whitespace.





\clearpage
\bibliographystyle{apalike}
\bibliography{references}

\clearpage
\section*{Supplementary Information}
\label{sec:appendix}
\setcounter{figure}{0}
\setcounter{table}{0}
\renewcommand{\thefigure}{S\arabic{figure}}
\renewcommand{\thetable}{S\arabic{table}}

\begin{table}[!ht]
\centering
\caption{\sc{Sample Summary Statistics}}
\label{tab:appendix-sample-summary}
\input{tables/table_appendix_sample_summary.tex}
\begin{minipage}{1\linewidth}
\footnotesize \textsl{Note.---} Panel A reports the city-level distribution of five variables across the 204-city sample. Panel B reports means by U.S.\ Census region. Hazard-relevant share is the share of a city's municipal code that GABRIEL classifies as hazard-relevant. Total NRI risk score is the sum across the 18 NRI hazards of the city's per-hazard risk score.
\end{minipage}
\end{table}

\begin{table}[!ht]
\centering
\caption{\sc{Region-Population-Weighted Robustness Check}}
\label{tab:appendix-region-weighted}
\input{tables/table_appendix_region_weighted.tex}
\begin{minipage}{1\linewidth}
\footnotesize \textsl{Note.---} I reweight each city so its Census region contributes to the sample in proportion to the region's share of the U.S. population (Northeast 17.2\%, Midwest 20.7\%, South 38.3\%, West 23.9\%; total weights sum to 204). Under this reweighting, each Northeast city receives a weight of 2.92, each Midwest city 1.28, each South city 0.90, and each West city 0.68.
\end{minipage}
\end{table}

\begin{table}[!ht]
\centering
\caption{\sc{Hazard-Relevant Share and Total NRI Risk}}
\label{tab:appendix-risk-regression}
\input{tables/table_appendix_risk_regression.tex}
\begin{minipage}{1\linewidth}
\footnotesize \textsl{Note.---} Each row is a city ($N = 204$). The dependent variable is the share of a city's municipal code that GABRIEL classifies as hazard-relevant, in percentage points. NRI risk score is the sum across the 18 NRI hazards of the city's per-hazard risk score, standardized across cities. Log(total sentences) is the natural log of the number of sentences in the city's code. Standard errors in parentheses. Significance: $^{*}p<0.10$, $^{**}p<0.05$, $^{***}p<0.01$.
\end{minipage}
\end{table}

\begin{table}[h!]
\centering
\caption{\textbf{Keywords used in the pre-filter, by hazard.} A sentence is sent to GABRIEL for scoring if it matches at least one term (case-insensitive). The ``General'' row contains broad hazard-related terms applied across all hazards.}
\label{tab:appendix-keywords}
\small
\begin{tabular}{p{0.18\textwidth} p{0.74\textwidth}}
\toprule
\textbf{Hazard} & \textbf{Keywords} \\
\midrule
Avalanche & avalanche \\
Coastal Flooding & storm surge; coastal flood; tidal flood; coastal erosion; coastal zone; coastal hazard; sea level; shore; shoreline; coastal inundation; coastal storm \\
Cold Wave & cold wave; extreme cold; freeze; freezing; frost; hypothermia; wind chill; cold emergency; cold weather \\
Drought & drought; water scarcity; water shortage; water conservation; water restriction; arid \\
Earthquake & earthquake; seismic; fault line; fault zone; liquefaction; richter; tremor; seismically \\
Hail & hail; hailstorm; hailstone \\
Heat Wave & heat wave; extreme heat; heat emergency; heat island; urban heat; heat stress; heat advisory; cooling center; excessive heat \\
Hurricane & hurricane; tropical storm; tropical cyclone; typhoon; tropical depression \\
Ice Storm & ice storm; ice accumulat; freezing rain; icing; ice load \\
Landslide & landslide; mudslide; mudflow; debris flow; slope stability; slope failure; mass movement; rockfall; rockslide \\
Lightning & lightning \\
Riverine Flooding & flood; floodplain; floodway; flood zone; flood hazard; flood insurance; flood damage; flood control; flood risk; flood elevation; flood level; inundation; levee; base flood; fema flood; 100-year; 100 year; stormwater; storm water; drainage; watershed \\
Strong Wind & wind speed; wind load; wind zone; high wind; strong wind; windstorm; wind resistant; wind exposure \\
Tornado & tornado; tornadic; safe room; storm shelter; storm cellar \\
Tsunami & tsunami \\
Volcanic Activity & volcano; volcanic; lava; eruption; lahar; pyroclastic; ashfall \\
Wildfire & wildfire; wildland; fire hazard; fire risk; fire zone; fire-resistant; fire resistant; fire protection; wui; urban interface; brush fire; brushfire; defensible space; fire break; firebreak \\
Winter Weather & snow load; snow removal; blizzard; winter storm; winter weather; ice dam; snow emergency \\
\midrule
General & hazard; disaster; emergency management; emergency preparedness; emergency plan; emergency response; evacuation; shelter in place; climate change; climate adaptation; climate resilience; resilience; resilient; mitigation; natural disaster; catastroph; fema; nfip; risk assessment; vulnerability; building code; international building code; ibc; erosion; subsidence; sinkhole; radon; endangered species; wetland; riparian; environmental protection; environmental hazard \\
\bottomrule
\end{tabular}
\end{table}

\begin{table}[h!]
\centering
\caption{\textbf{Top 25 external regulatory frameworks referenced in municipal codes.} Frameworks are ranked by the number of cities in the 204-city sample that contain at least one reference. ``Named'' entries are full framework names; ``acronym'' entries are abbreviations. The full dictionary of 384 entries (142 named frameworks and 242 acronyms) is provided as a supplementary data file.}
\label{tab:appendix-external-frameworks}
\input{tables/table_external_frameworks.tex}
\end{table}

\input{tables/table_gabriel_prompts.tex}

\begin{table}[!ht]
\centering
\caption{\sc{Classifier Validation Against Expert-Coded Ground Truth}}
\label{tab:appendix-validation}
\input{tables/table_appendix_validation.tex}
\begin{minipage}{1\linewidth}
\footnotesize \textsl{Note.---} Panel I reports validation on the 210 sentences that GABRIEL rated (a stratified random sample: 148 that GABRIEL scored as hazard-relevant, plus 62 that GABRIEL scored as not hazard-relevant). Each sentence was coded by hand, without reference to the model's score. The first row reports the overall binary classification of a sentence as hazard-relevant; the subsequent rows report each of the 18 hazards, with $N$ giving the number of sentences the human called positive for that hazard. Panel II reports the share of a separate random sample of 89 sentences that were dropped by the keyword pre-filter (and never seen by GABRIEL) that the human coder called hazard-relevant, with a Wilson 95\% confidence interval.
\end{minipage}
\end{table}

\begin{table}[!ht]
\centering
\caption{\sc{Main Results at Alternative Population Thresholds}}
\label{tab:appendix-threshold}
\input{tables/table_appendix_threshold.tex}
\begin{minipage}{1\linewidth}
\footnotesize \textsl{Note.---} Each row restricts the sample to cities at or above the population threshold and recomputes the paper's three main quantities on that subsample. Hazard-relevant share is the median across cities of the share of code sentences classified as hazard-relevant. Riverine flood share is the mean across cities of the share of hazard-relevant text addressing riverine flooding. The text-risk gap is the median across cities. The first row is the sample used in the paper.
\end{minipage}
\end{table}

\begin{table}[!ht]
\centering
\caption{\sc{Vintage of the Code Sections Containing Hazard Provisions}}
\label{tab:appendix-code-vintage}
\resizebox{\textwidth}{!}{\input{tables/table_appendix_code_vintage.tex}}
\begin{minipage}{1\linewidth}
\footnotesize \textsl{Note.---} Each code section ends with a list of the ordinances that created and amended it. I process the raw section files, extract every year these lists contain, and use the earliest as the date of first adoption and the most recent as the date of last amendment. Each hazard-relevant sentence inherits the dates of the section it was drawn from. $N$ is the number of hazard-relevant sentences assigned to that hazard for which a date could be recovered; a sentence scoring above the cutoff on more than one hazard is counted under each. The final row covers all 39{,}642 datable hazard-relevant sentences, including those that are hazard-relevant on the overall score without exceeding the cutoff on any single hazard. Avalanche, hail, and volcanic activity have fewer than 30 datable sentences and are omitted.
\end{minipage}
\end{table}

\begin{figure}[h!]
\centering
\includegraphics[width=\textwidth]{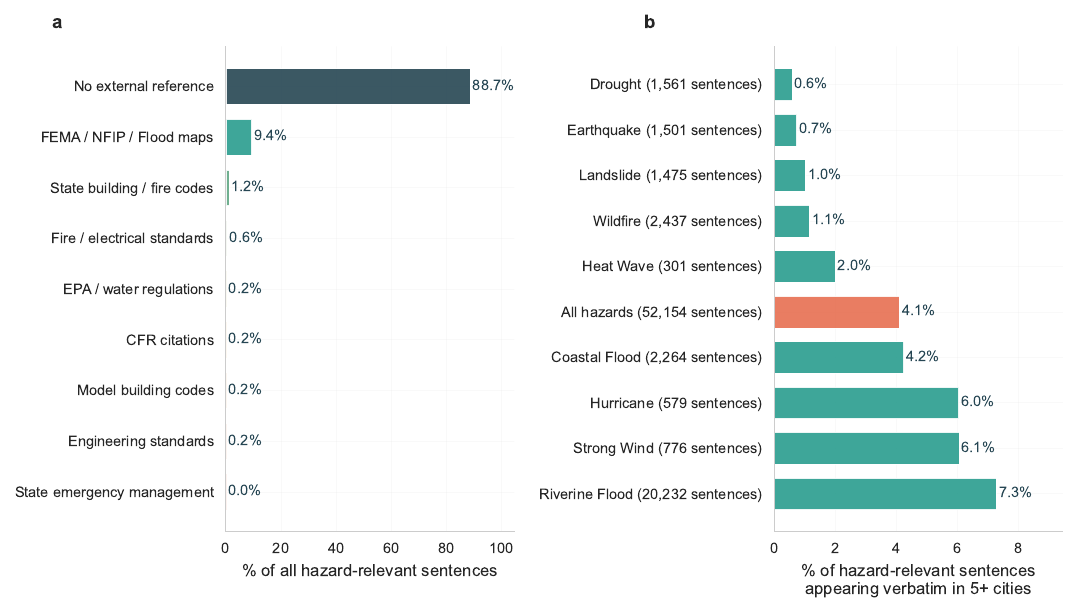}
\caption{\textbf{External framework references and boilerplate in hazard-relevant text.} (a) Share of all hazard-relevant sentences across the sample, classified by whether the sentence references an external regulatory framework. ``No external reference'' includes sentences without any reference to an external framework. The remaining sentences are grouped by the type of framework referenced (FEMA/NFIP, model building codes, state building or fire codes, and so on). (b) Share of hazard-relevant sentences classified as boilerplate. The highlighted bar shows the share across all hazards combined; the remaining bars show the share for each hazard individually. A sentence is classified as boilerplate if it appears verbatim in five or more cities. Numbers in parenthesis give the total count of hazard-relevant sentences in that category.}
\label{fig:appendix-local-boilerplate}
\end{figure}

\begin{figure}[h!]
\centering
\includegraphics[width=0.7\textwidth]{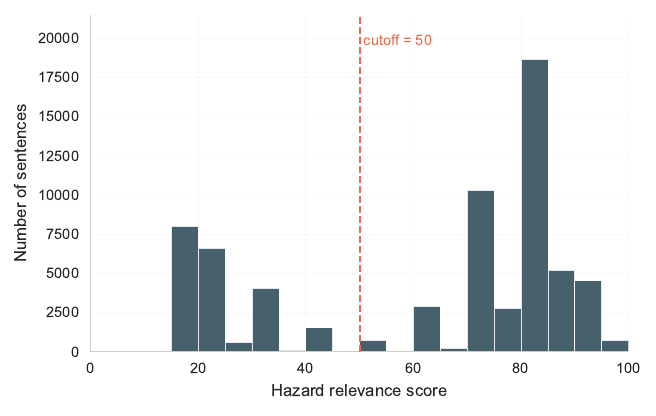}
\caption{\textbf{Distribution of hazard relevance scores assigned by GABRIEL.} Histogram of Stage 1 hazard-relevance scores for the 300{,}979 sentences that pass the pre-filter and are scored by GABRIEL. Sentences that did not pass the pre-filter are assigned a score of zero and are not shown. The dashed line at 50 marks the cutoff used to classify a sentence as hazard-relevant.}
\label{fig:appendix-relevance-distribution}
\end{figure}

\begin{figure}[h!]
\centering
\includegraphics[width=\textwidth]{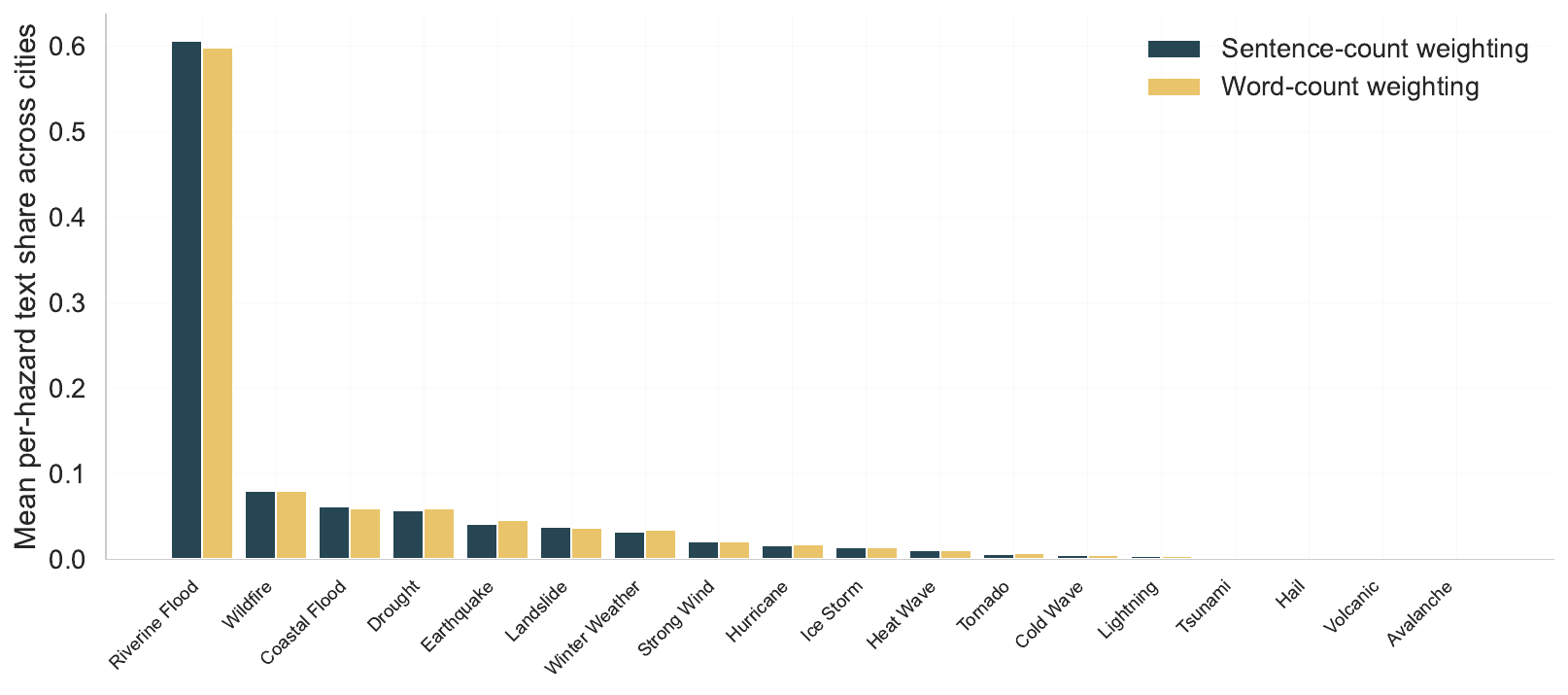}
\caption{\textbf{Word-count-weighted robustness check for the sentence-level classification.} Mean per-hazard text share across the 204 cities under two weightings. Sentence-count weighting (blue): each hazard-relevant sentence contributes one unit. Word-count weighting (yellow): each sentence contributes its word count.}
\label{fig:appendix-wordweight}
\end{figure}

\begin{figure}[h!]
\centering
\includegraphics[width=0.62\textwidth]{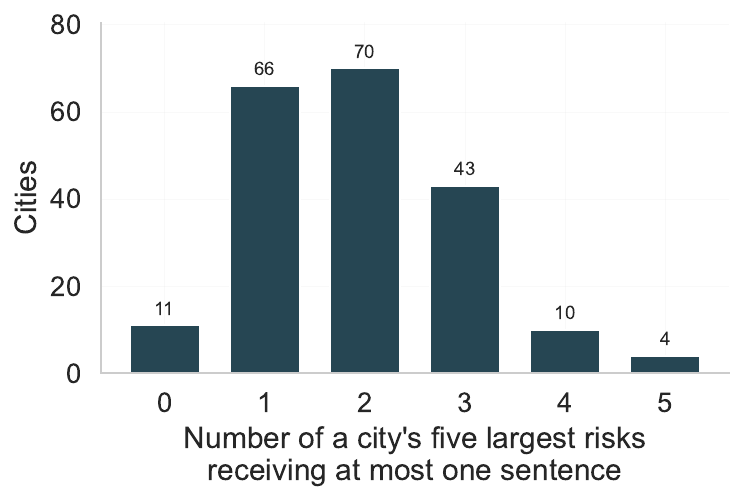}
\caption{\textbf{Text devoted to the hazards that pose a city's largest risks.} Hazards are ranked within each city by their share of that city's total NRI risk. A hazard is recorded as receiving the equivalent of at most one sentence, counting sentences fractionally where one addresses several hazards. The bars give the distribution across the 204 cities of how many of a city's five largest risks fall at or below that level.}
\label{fig:appendix-silence}
\end{figure}

\begin{figure}[h!]
\centering
\includegraphics[width=\textwidth]{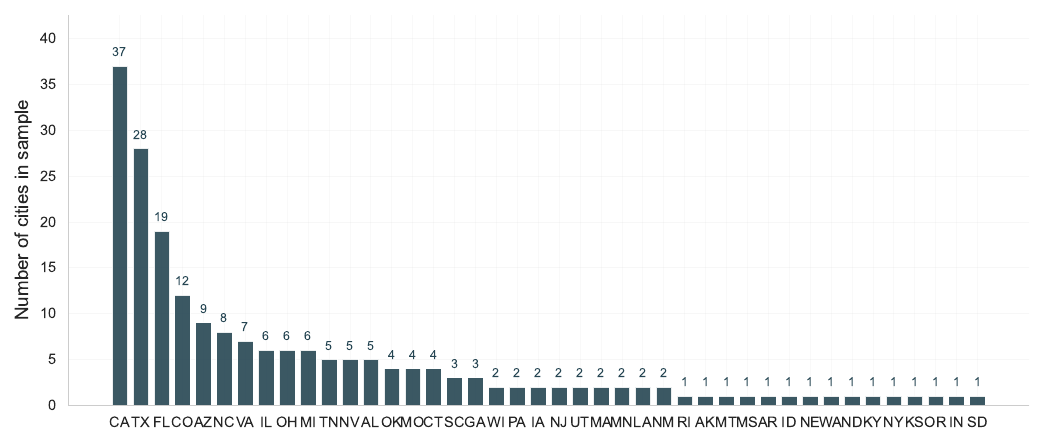}
\caption{\textbf{Cities per state in the sample.} Each bar shows the number of cities from a given state, sorted from most to fewest. The sample includes cities from 42 of 50 U.S. states.}
\label{fig:appendix-state-coverage}
\end{figure}

\begin{figure}[h!]
\centering
\includegraphics[width=0.7\textwidth]{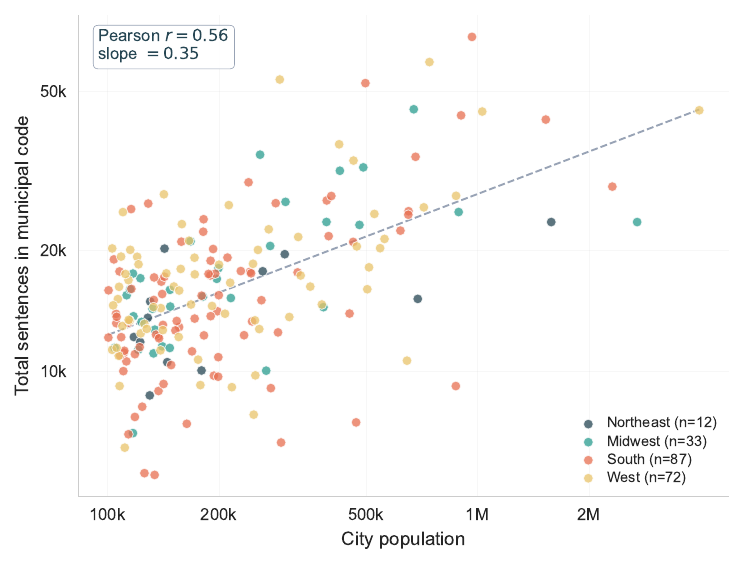}
\caption{\textbf{Population and code length across cities.} Each dot is one city, positioned by its population and by the total number of sentences in its municipal code. Both axes are on a log scale. Color indicates U.S.\ Census region. The dashed line is the OLS fit through the log-log points. Population and code length are moderately correlated (Pearson $r=0.54$).}
\label{fig:appendix-pop-code-length}
\end{figure}

\end{document}

%% file: tables/table_appendix_sample_summary.tex
\resizebox{1\textwidth}{!}{\begin{tabular}{lccccc}
\toprule\toprule
& Mean & SD & Min & Median & Max \\
& \multicolumn{5}{c}{\sc{Panel A. City-level distribution (N = 204)}} \\ \cmidrule(r){2-6}
Population & 297,011 & 409,938 & 100,509 & 171,892 & 3,973,278 \\
Total code length (sentences) & 17,779 & 9,389 & 5,504 & 15,464 & 68,492 \\
Hazard-relevant sentences & 256 & 126 & 28 & 247 & 1,149 \\
Hazard-relevant share (\%) & 1.54 & 0.59 & 0.41 & 1.55 & 3.54 \\
Total NRI risk score & 582.40 & 143.43 & 273.31 & 591.88 & 922.56 \\
\bottomrule
\end{tabular}}

\vspace{0.4cm}

\resizebox{1\textwidth}{!}{\begin{tabular}{lcccc}
\toprule\toprule
& N cities & Mean population & Mean code length & Mean hazard share (\%) \\
& \multicolumn{4}{c}{\sc{Panel B. Means by U.S. Census region}} \\ \cmidrule(r){2-5}
Northeast & 12 & 327,594 & 14,849 & 1.06 \\
Midwest & 33 & 319,061 & 18,550 & 1.40 \\
South & 87 & 275,959 & 17,070 & 1.67 \\
West & 72 & 307,245 & 18,771 & 1.52 \\
\bottomrule
\end{tabular}}

%% file: tables/table_appendix_region_weighted.tex
\resizebox{1\textwidth}{!}{\begin{tabular}{lrrr}
\toprule\toprule
& Unweighted & Region-pop-weighted & $\Delta$ \\
\midrule
Mean hazard-relevant share & 1.54\% & 1.48\% & -0.064 \\
Median text-risk gap & 0.697 & 0.698 & +0.000 \\
Mean text-risk gap & 0.684 & 0.690 & +0.006 \\
Riverine flooding share of hazard text & 60.67\% & 60.58\% & -0.093 \\
Wildfire share of hazard text & 7.98\% & 7.45\% & -0.529 \\
Coastal flood share of hazard text & 6.25\% & 6.54\% & +0.288 \\
Drought share of hazard text & 5.74\% & 5.31\% & -0.425 \\
Earthquake share of hazard text & 4.14\% & 3.30\% & -0.844 \\
Winter weather share of hazard text & 3.25\% & 4.59\% & +1.341 \\
\bottomrule
\end{tabular}}

%% file: tables/table_appendix_risk_regression.tex
\resizebox{1\textwidth}{!}{\begin{tabular}{lcc}
\toprule\toprule
& (I) & (II) \\
& Baseline & With log(code length) \\
& \multicolumn{2}{c}{\sc{Dependent variable: hazard-relevant share (percentage points)}} \\ \cmidrule(r){2-3}
NRI risk score ($z$-scored)   & +0.145$^{***}$ & +0.103$^{***}$ \\
                              & (0.040) & (0.038) \\
Log(total sentences)          &         & -0.476$^{***}$ \\
                              &         & (0.087) \\
Constant                      & +1.541$^{***}$ & +6.147$^{***}$ \\
                              & (0.040) & (0.848) \\
\midrule
Observations                  & 204 & 204 \\
$R^{2}$                     & 0.061 & 0.181 \\
\bottomrule
\end{tabular}}

%% file: tables/table_external_frameworks.tex
\small
\begin{tabular}{l l r r}
\toprule
\textbf{Framework} & \textbf{Type} & \textbf{\# cities} & \textbf{\# mentions} \\
\midrule
Building Code & named & 181 & 2730 \\
Federal Emergency Management Agency & named & 180 & 1074 \\
National Electrical Code & named & 178 & 2036 \\
Federal Water Pollution Control Act & named & 171 & 379 \\
EPA & acronym & 168 & 737 \\
FEMA & acronym & 163 & 974 \\
FIRM & acronym & 160 & 928 \\
National Flood Insurance Program & named & 157 & 584 \\
National Fire Protection Association & named & 155 & 438 \\
Fire Code & named & 149 & 1775 \\
International Fire Code & named & 141 & 3634 \\
International Building Code & named & 140 & 4441 \\
AO & acronym & 137 & 386 \\
NFPA & acronym & 129 & 829 \\
International Residential Code & named & 125 & 1922 \\
FIS & acronym & 110 & 260 \\
Electrical Code & named & 110 & 632 \\
Mechanical Code & named & 110 & 372 \\
Plumbing Code & named & 109 & 580 \\
National Electrical Safety Code & named & 104 & 220 \\
International Existing Building Code & named & 101 & 469 \\
International Mechanical Code & named & 98 & 1222 \\
International Plumbing Code & named & 96 & 1319 \\
Uniform Building Code & named & 82 & 1242 \\
BFE & acronym & 81 & 204 \\
\bottomrule
\end{tabular}

%% file: tables/table_gabriel_prompts.tex

\begin{table}[h!]
\centering
\caption{\textbf{Attribute definitions used by GABRIEL for sentence classification.} Each row shows the attribute name and the exact text description passed to the underlying language model (GPT-4o-mini) at scoring time. Classification is performed in two stages: Stage 1 scores every sentence that passes the keyword pre-filter (Table~\ref{tab:appendix-keywords}) on \texttt{hazard\_relevance} and \texttt{enforceability}; Stage 2 scores sentences with Stage 1 \texttt{hazard\_relevance} $\geq$ 15 on the remaining attributes. All calls use temperature $= 0$ and structured JSON output.}
\label{tab:appendix-gabriel-prompts}
\footnotesize
\begin{tabular}{p{0.20\textwidth} p{0.72\textwidth}}
\toprule
\textbf{Attribute} & \textbf{Definition passed to the model} \\
\midrule
\multicolumn{2}{l}{\textit{Panel A: Overall hazard relevance and legal enforceability (Stage 1)}} \\
\midrule
hazard\_relevance & How directly this sentence addresses natural hazard risk, climate adaptation, or disaster preparedness. 100 = explicitly about hazard risk, climate resilience, or emergency management. 0 = unrelated municipal business (zoning setbacks, parking requirements, signage rules, etc.). \\
\Tnew
enforceability & How enforceable and legally binding this sentence is. 100 = mandatory prescriptive language (shall, must, required, prohibited). 0 = aspirational advisory language (should, may, encourage, recommend). \\
\midrule
\multicolumn{2}{l}{\textit{Panel B: The 18 hazard-specific attributes from FEMA's National Risk Index (Stage 2)}} \\
\midrule
avalanche & How directly this sentence addresses avalanche risk, avalanche zones, or avalanche mitigation. 100 = explicitly about avalanches. 0 = not about avalanches. \\
\T
coastal\_flooding & How directly this sentence addresses coastal flooding, storm surge, coastal flood zones, or tidal flooding. 100 = explicitly about coastal flooding. 0 = not about coastal flooding. \\
\T
cold\_wave & How directly this sentence addresses cold waves, extreme cold events, or cold weather emergencies. 100 = explicitly about cold waves. 0 = not about cold waves. \\
\T
drought & How directly this sentence addresses drought, water scarcity, or drought preparedness. 100 = explicitly about drought. 0 = not about drought. \\
\T
earthquake & How directly this sentence addresses earthquakes, seismic risk, or earthquake-resistant construction. 100 = explicitly about earthquakes. 0 = not about earthquakes. \\
\T
hail & How directly this sentence addresses hail, hailstorms, or hail damage mitigation. 100 = explicitly about hail. 0 = not about hail. \\
\T
heat\_wave & How directly this sentence addresses heat waves, extreme heat events, heat emergencies, or urban heat. 100 = explicitly about heat waves. 0 = not about heat waves. \\
\T
hurricane & How directly this sentence addresses hurricanes, tropical storms, or hurricane preparedness. 100 = explicitly about hurricanes. 0 = not about hurricanes. \\
\T
ice\_storm & How directly this sentence addresses ice storms, freezing rain, or ice accumulation hazards. 100 = explicitly about ice storms. 0 = not about ice storms. \\
\T
landslide & How directly this sentence addresses landslides, mudslides, slope stability, or debris flows. 100 = explicitly about landslides. 0 = not about landslides. \\
\T
lightning & How directly this sentence addresses lightning strikes, lightning protection, or thunderstorm electrical hazards. 100 = explicitly about lightning. 0 = not about lightning. \\
\T
riverine\_flooding & How directly this sentence addresses river flooding, floodplains, flood zones, or inland flooding. 100 = explicitly about riverine flooding. 0 = not about riverine flooding. \\
\T
strong\_wind & How directly this sentence addresses strong winds, wind load requirements, or high-wind hazards (excluding hurricanes and tornadoes). 100 = explicitly about strong winds. 0 = not about strong winds. \\
\T
tornado & How directly this sentence addresses tornadoes, tornado shelters, or tornado preparedness. 100 = explicitly about tornadoes. 0 = not about tornadoes. \\
\T
tsunami & How directly this sentence addresses tsunamis, tsunami evacuation zones, or tsunami preparedness. 100 = explicitly about tsunamis. 0 = not about tsunamis. \\
\T
volcanic\_activity & How directly this sentence addresses volcanic activity, volcanic hazard zones, or volcanic eruption preparedness. 100 = explicitly about volcanic activity. 0 = not about volcanic activity. \\
\T
wildfire & How directly this sentence addresses wildfires, wildfire risk zones, fire-resistant construction, or wildland-urban interface. 100 = explicitly about wildfires. 0 = not about wildfires. \\
\T
winter\_weather & How directly this sentence addresses winter weather, snow loads, blizzards, or winter storm preparedness. 100 = explicitly about winter weather. 0 = not about winter weather. \\
\midrule
\multicolumn{2}{l}{\textit{Panel C: Regulatory mechanism classification (Stage 2, fixed-scale categorical)}} \\
\midrule
regulatory\_mechanism & What type of regulatory mechanism this sentence represents. Score the SINGLE best match on this scale: 0 = not a regulation. 20 = zoning or land use (setbacks, permitted uses, overlay districts, density). 40 = building code or construction standard (structural, materials, design). 60 = environmental regulation (stormwater, wetlands, tree canopy, erosion control). 80 = infrastructure or public works (drainage, utilities, roads, levees). 100 = emergency management (evacuation, shelters, disaster plans, sirens). \\
\bottomrule
\end{tabular}
\end{table}

%% file: tables/table_appendix_validation.tex
\resizebox{1\textwidth}{!}{\begin{tabular}{lrrrr}
\toprule\toprule
\multicolumn{5}{l}{\sc{Panel I. Sentences GABRIEL rated}} \\
\midrule
& N & Precision & Recall & F1 \\
\midrule
Sentence is hazard-relevant   & 210 & 0.88 & 0.86 & 0.87 \\
\midrule
Per-hazard (18-way, multi-label) &  &  &  &  \\
\midrule
Riverine Flood & 46 & 1.00 & 0.33 & 0.49 \\
Earthquake & 21 & 0.95 & 0.90 & 0.93 \\
Winter Weather & 20 & 0.62 & 0.90 & 0.73 \\
Strong Wind & 17 & 0.73 & 0.65 & 0.69 \\
Hurricane & 16 & 0.84 & 1.00 & 0.91 \\
Coastal Flood & 11 & 0.50 & 0.73 & 0.59 \\
Wildfire & 9 & 0.64 & 0.78 & 0.70 \\
Landslide & 9 & 0.60 & 0.67 & 0.63 \\
Cold Wave & 9 & 0.54 & 0.78 & 0.64 \\
Hail & 8 & 1.00 & 1.00 & 1.00 \\
Drought & 8 & 0.60 & 0.38 & 0.46 \\
Tornado & 8 & 0.75 & 0.75 & 0.75 \\
Volcanic & 8 & 0.80 & 1.00 & 0.89 \\
Avalanche & 6 & 0.86 & 1.00 & 0.92 \\
Heat Wave & 5 & 0.29 & 0.40 & 0.33 \\
Tsunami & 5 & 0.45 & 1.00 & 0.62 \\
Lightning & 4 & 0.33 & 1.00 & 0.50 \\
Ice Storm & 0 & — & — & — \\
\midrule
\multicolumn{5}{l}{\sc{Panel II. Share of pre-filter-dropped sentences that were hazard-relevant}} \\
\midrule
& Sampled & Hazard-relevant & Share & 95\\
\midrule
Dropped by keyword filter      & 89 & 2 & 2.2\% & [0.6\%, 7.8\%] \\
\bottomrule
\end{tabular}}

%% file: tables/table_appendix_threshold.tex
\begin{tabular}{lcccc}
\toprule\toprule
Population threshold & Cities & Hazard-relevant & Riverine flood & Text-risk \\
 & & share (\%) & share (\%) & gap \\
\midrule
100,000 (paper) & 204 & 1.55 & 60.7 & 0.70 \\
150,000 & 117 & 1.26 & 59.6 & 0.68 \\
200,000 & 78 & 1.16 & 56.7 & 0.67 \\
250,000 & 63 & 1.14 & 55.5 & 0.66 \\
\bottomrule
\end{tabular}

%% file: tables/table_appendix_code_vintage.tex
\begin{tabular}{lrrrrrr}
\toprule\toprule
& & \multicolumn{2}{c}{First ordinance} & \multicolumn{2}{c}{Last amendment} & \\
\cmidrule(lr){3-4}\cmidrule(lr){5-6}
Hazard & $N$ & Median & 10th pct. & Median & 10th pct. & \% pre-2000 \\
\midrule
Riverine Flood & 15,492 & 2004 & 1967 & 2020 & 2011 & 3\% \\
Coastal Flood & 1,717 & 2000 & 1967 & 2020 & 2014 & 3\% \\
Wildfire & 1,624 & 1996 & 1956 & 2020 & 2009 & 5\% \\
Drought & 1,194 & 1994 & 1952 & 2020 & 2014 & 2\% \\
Landslide & 929 & 1997 & 1966 & 2020 & 2008 & 6\% \\
Earthquake & 884 & 1997 & 1956 & 2020 & 2008 & 7\% \\
Winter Weather & 657 & 1993 & 1961 & 2020 & 2009 & 5\% \\
Strong Wind & 576 & 1996 & 1966 & 2020 & 2013 & 3\% \\
Hurricane & 503 & 1996 & 1964 & 2020 & 2011 & 2\% \\
Ice Storm & 360 & 1993 & 1961 & 2019 & 2009 & 7\% \\
Heat Wave & 204 & 1996 & 1961 & 2020 & 2012 & 4\% \\
Cold Wave & 147 & 1994 & 1961 & 2019 & 2006 & 7\% \\
Tornado & 122 & 1997 & 1964 & 2019 & 2009 & 2\% \\
Lightning & 109 & 1993 & 1953 & 2019 & 2006 & 5\% \\
Tsunami & 41 & 1996 & 1967 & 2020 & 2010 & 7\% \\
\midrule
All hazard-relevant text & 39,642 & 1998 & 1962 & 2020 & 2010 & 4\% \\
\bottomrule
\end{tabular}